\documentclass[a4paper,12pt]{article}
\usepackage[utf8]{inputenc}
\usepackage{amsmath}
\usepackage{amssymb}
\usepackage{bbold}
\usepackage{amsfonts,bm}
\usepackage{geometry}
\usepackage{fullpage}
\usepackage{graphicx}
\usepackage{caption}
\usepackage{todonotes}
\usepackage{multicol}
\usepackage{subcaption}
\usepackage{slashed}
\usepackage{float}
\usepackage{color,xcolor}
\usepackage{epstopdf}
\usepackage{units}
\usepackage[normalem]{ulem} 
\usepackage{jheppub} 
\usepackage{physics}
\usepackage{bm,paralist,xspace,url,relsize}
\usepackage[]{hyperref}
\usepackage{isotope}

\def\Emax{$E_{\rm max}$}
\newcommand{\Tl}{\isotope[204]{Tl}}
\newcommand{\Tm}{\isotope[171]{Tm}}
\newcommand{\Ba}{\isotope[133]{Ba}}
\newcommand{\Co}{\isotope[60]{Co}}

\title{Measurement of the \Tm\ beta spectrum}

 \author[a]{Fr\'{e}d\'{e}ric~Juget,}
 \author[b]{Maarten~van~Dijk,}
 \author[c]{Emilio~Andrea~Maugeri,}
 \author[c]{Maria~Dorothea~Schumann,}
 \author[d]{Stephan~Heinitz,}
 \author[b, e]{Alexey~Boyarsky,}
 \author[f]{Ulli~K\"{o}ster,}
 \author[b]{Lesya~Shchutska,}
 \author[a]{Claude~Bailat}
 \affiliation[a]{Institute of Radiation Physics, Lausanne, Switzerland}
 \affiliation[b]{Institute of Physics,
École Polytechnique Fédérale de Lausanne (EPFL), Lausanne, Switzerland}
 \affiliation[c]{Paul Scherrer Institute, Villigen, Switzerland}
  \affiliation[d]{Belgian Nuclear Research Center, Mol, Belgium}
  \affiliation[e]{Instituut-Lorentz, Leiden University, Leiden, The Netherlands}
\affiliation[f]{Institut Laue-Langevin, Grenoble, France}
\emailAdd{Frederic.Juget@chuv.ch}
\date{}

\begin{document}
\abstract{The beta spectrum of the main transition of the \Tm\ was measured using a double focalizing spectrometer. The instrument was lately improved in order to reduce its low energy threshold to 34 keV. We used the spectrometer to measure the energy end-point of the main transition of \Tm\ using the Kurie plot formalism. We report a new value of 97.60(38) keV, which is in agreement with previous measurements. In addition, the spectrum shape was compared with the $\xi$-approximation calculation where the shape factor is equal to 1 and good agreement was found between the theory and the measurement at the 1\% level.}
\keywords{\Tm, Beta spectrum measurement, Magnetic spectrometer, Kurie plot, End-point energy}
\maketitle

\section{Introduction}
\label{sec:introduction}
\begin{figure}[!h]
    \centering
    \includegraphics[width=0.7\textwidth]{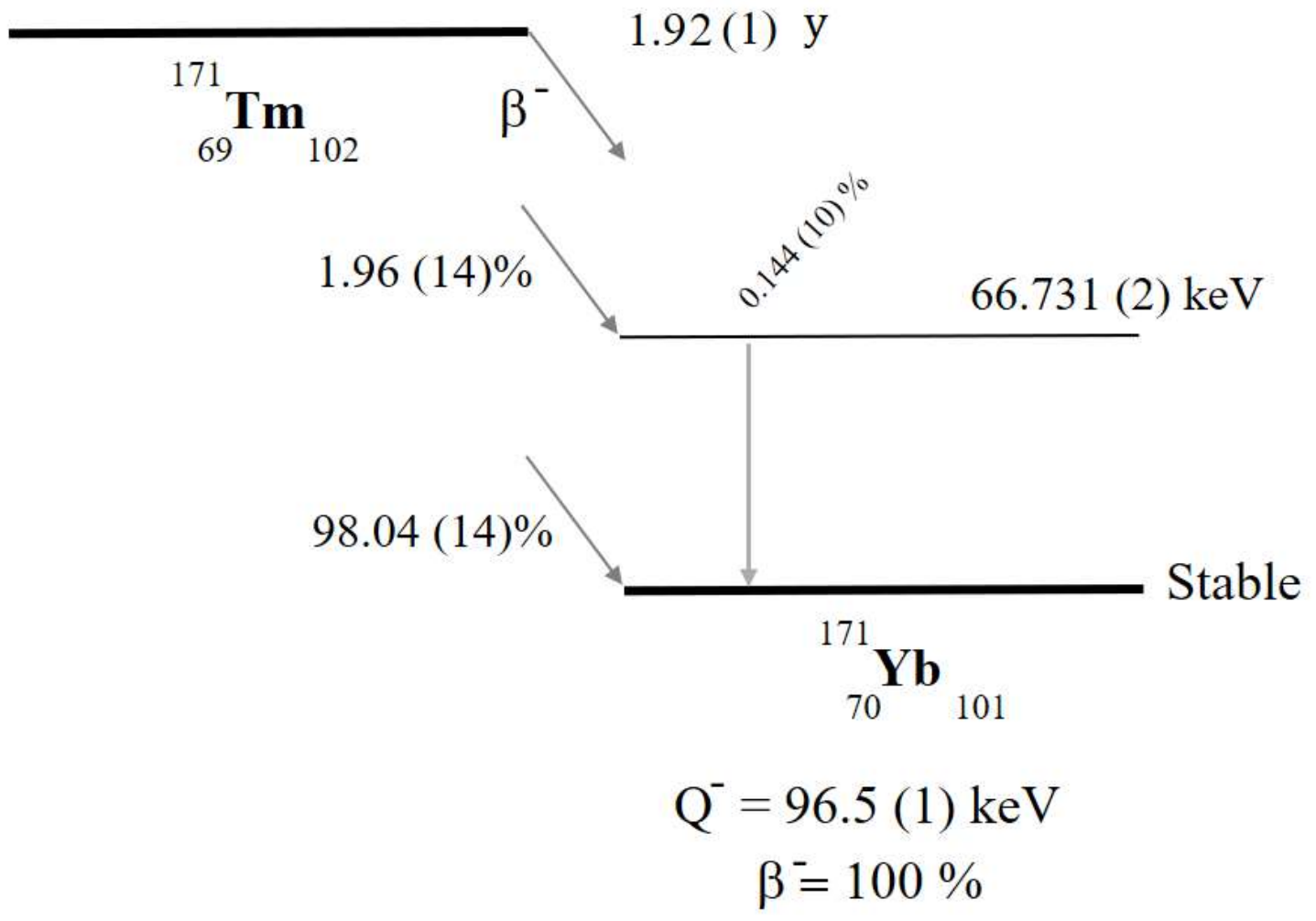}
    \caption{Decay scheme of \Tm\ from ENSDF~\cite{ENSDF}.}
    \label{fig:fig1}
\end{figure}

\begin{figure}[!ht]
    \centering
    \includegraphics[width=0.9\textwidth]{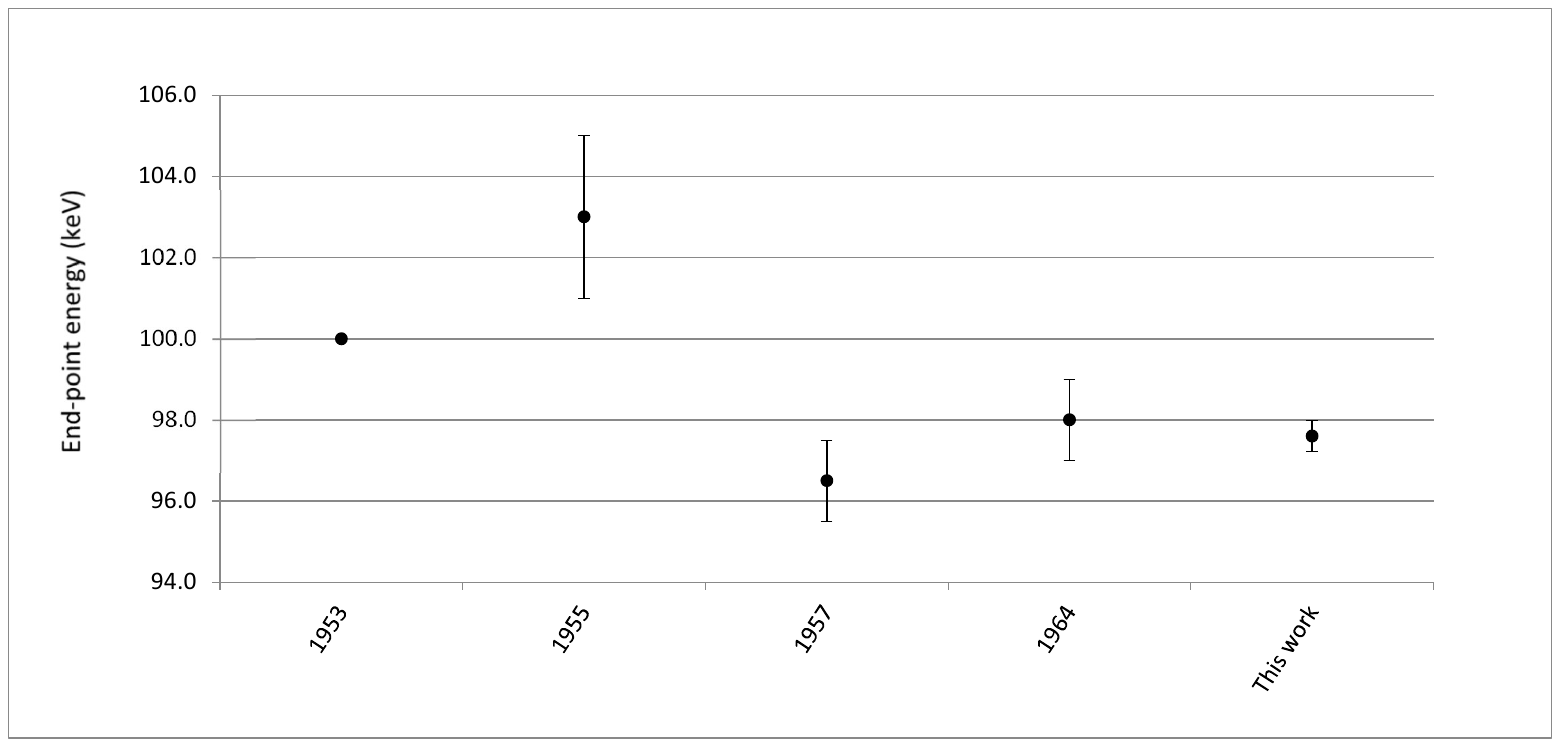}
    \caption{End-point energy values from previous measurements in chronological order, 1953~\cite{Hollander1953-zr}; 100 keV, 1955~\cite{Bisi}; 103(2) keV, 1957~\cite{Smith:1957zz}; 96.5(10) keV, 1964~\cite{Hansen}; 98(1) keV, with the final value obtained in this work}
    \label{fig:fig2}
\end{figure}

Since one decade the interest in measuring beta decay spectra has increased, in particular in radionuclide metrology~\cite{Kossert:2018ydo,Kossert:2015ydo}, astrophysics~\cite{Brdar:2022wuv}, nuclear physics~\cite{Mention:2011ydo} and in nuclear medicine where more and more beta emitters for targeted radionuclide therapy are being studied ~\cite{Qaim:2018,Naskar:2021,IAEA:2021}. Following our measurements of the shape of beta spectra using a double focalizing spectrometer~\cite{Juget2014-il,Juget2019-fj}, several improvements were performed on our facility. We were able to reduce the low energy threshold and increase the energy resolution. Using the modified system, the ground state decay beta spectrum of \Tm\ was measured. This isotope has two first-forbidden non-unique decay transitions with the dominant branch leading to the ground state (Fig.~\ref{fig:fig1}). Currently, only the decay of excited state of \isotope[171]{Yb} was measured using its coincidence with the de-excitation gamma ray line~\cite{Smith:1957zz,Robinson,Hansen}. These results reported a compatibility with an allowed transition shape but due to large uncertainties of the experimental setup, the authors could not conclude that the spectrum had an allowed shape~\cite{Robinson}. A recent and more detailed study~\cite{Mikulenko:2021ydo,Brdar:2022wuv}, showed that as \Tm\ has a low Q value and its $\xi$ value is large, where $\xi$ = Z$\alpha$/QR ($\alpha$ is the fine structure constant and R = (1.2 fm)A$^{1/3}$ is a typical nuclear radius), the decay spectrum shape of both transitions should be compatible with an allowed shape up to O(1\%) corrections. They revised the data from Ref.~\cite{Robinson} to obtain the shape factor parametrizations according to a first-forbidden non-unique transition. They concluded that ``none of the parametrizations provide a markedly better fit than a flat line (the allowed approximation)''. The interpretation is that in absence of precise data, the validity of the $\xi$-approximation can be supported at the level of 10\% for the secondary branch. In addition, \Tm\ is also an attractive candidate for the detection of the cosmic neutrino background as proposed by the PTOLEMY collaboration~\cite{PTOLEMY:2019hkd}. The \Tm\ neutrino capture cross section depends on the spectrum shape and using the $\xi$-approximation provides a prediction at 1\% level of the cross section, which is a necessary, although not sufficient condition for the use of \Tm\ for the cosmic neutrino background measurement~\cite{Brdar:2022wuv}.  End-point energy measurements of the main transition were performed in the 1950’s~\cite{Smith:1957zz,Robinson,Bisi} and 1960’s~\cite{Hansen}, which suffer from large discrepancies (Fig.~\ref{fig:fig2}). In addition, no detailed uncertainty calculations are reported. In light of the previous studies, there is a clear need for new measurements, especially for the ground state decay spectrum which has never been measured yet, in order to reach a definitive conclusion on the $\xi$-approximation and obtain the end-point energy value of the ground state decay.

 In this work, the calibration of the improved spectrometer is presented in Section~\ref{sec:spectrometer} as well as the energy resolution measured with a \Ba\ source and the calculation of the efficiency using a \Co\ source. Section~\ref{sec:source} reports the \Tm\ source preparation and Section~\ref{sec:measurement} shows the results of the \Tm\ measurement. 


\section{Measurement with the double focalizing magnetic spectrometer}
\label{sec:spectrometer}
The principle of operation of the double focalizing magnetic spectrometer is described in Refs.~\cite{Sakai1960-gk,Siegbahn1966-kr}. It consists of focalizing the electrons from a source toward a detector in both horizontal and vertical planes using a horizontal deflection angle of 180$^\circ$ which selects the energy and optimizes the detector collection efficiency. The magnetic spectrometer used in this work is described in Refs.~\cite{Juget2014-il,Juget2019-fj}. For the measurement of \Tm\ spectrum, a silicon detector with an active thickness of 500 $\mu$m was used in order to reach an energy threshold of 34 keV, imposed by the electronic noise, and to improve the energy resolution in comparison to previous reported measurements using a thicker detector~\cite{Juget2014-il,Juget2019-fj}. A LabView acquisition system records the amplitude of the pulse height in the detector, which corresponds to the energy deposited by the electron, using a National Instrument PCI-6115 DAQ board. The offline analysis was performed using the ROOT framework~\cite{Brun:1997pa}. The analysis uses a window around the full energy deposition peak to select only electrons which deposited all kinetic energy in the detector~\cite{Juget2019-fj}. To reduce the background and the backscattering in material around the source location, a 5\,mm thick Plexiglas cylinder is placed as shielding around the source area with a hole of 1 cm diameter in front of the source. 

\subsection{Energy calibration and resolution}
\label{sec:calibration}
\begin{figure}[!h]
    \centering
    \includegraphics[width=0.49\textwidth]{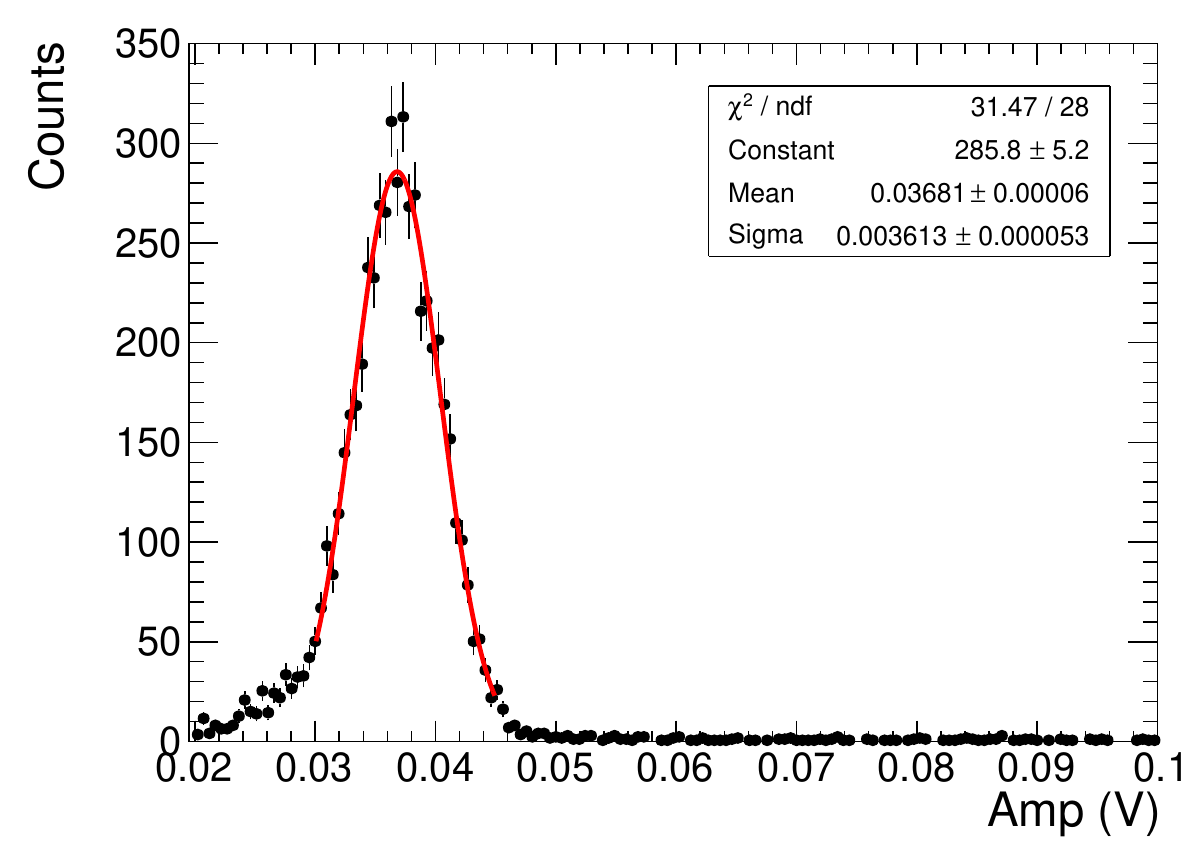}
    \includegraphics[width=0.49\textwidth]{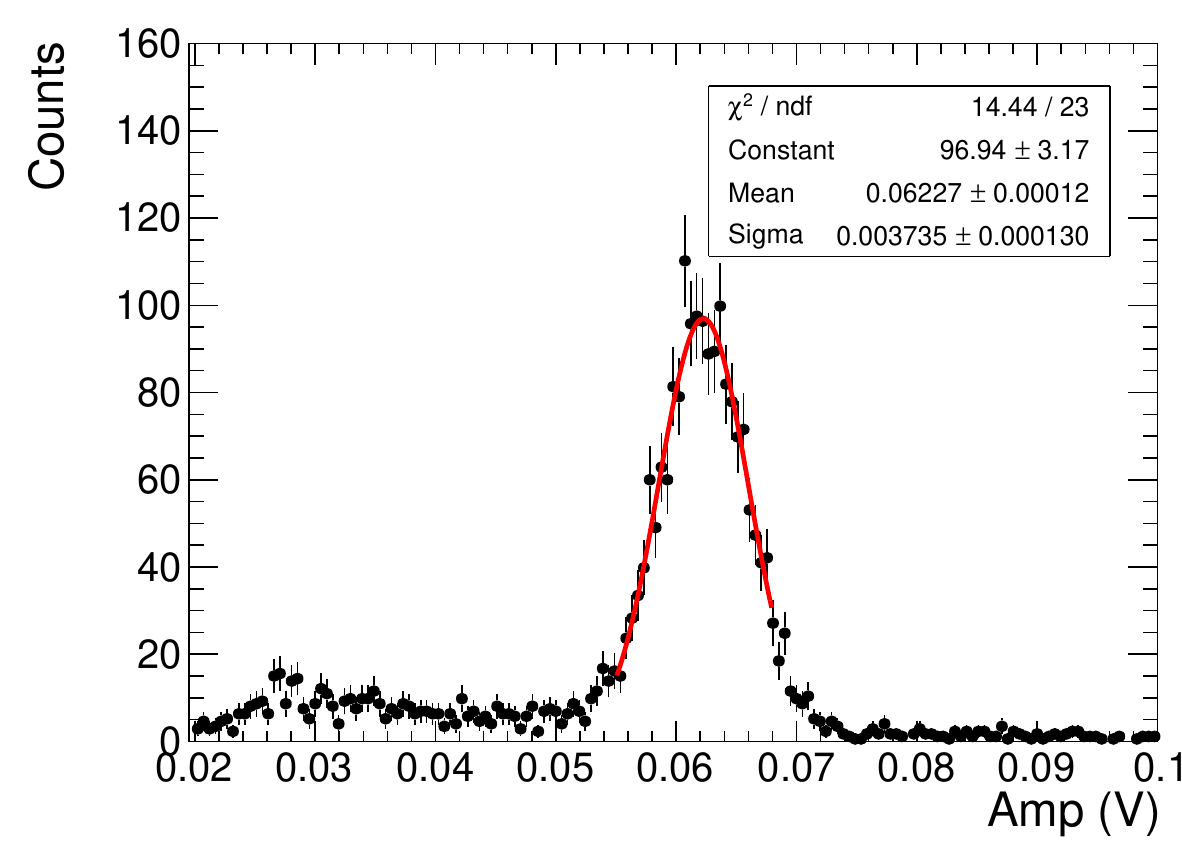}
    \caption{Gaussian fit on the distribution of the amplitude of the pulse height (Amp) for (left) 45.01\,keV and (right) 75.38\,keV electrons of a \Ba\ source measured for a given fixed value of the magnetic field after background subtraction. The mean value corresponds to the energy of the conversion electron at 45.01\,keV and 75.38\,keV.}
    \label{fig:fig3}
\end{figure}

\begin{figure}[!h]
    \centering
    \includegraphics[width=0.65\textwidth]{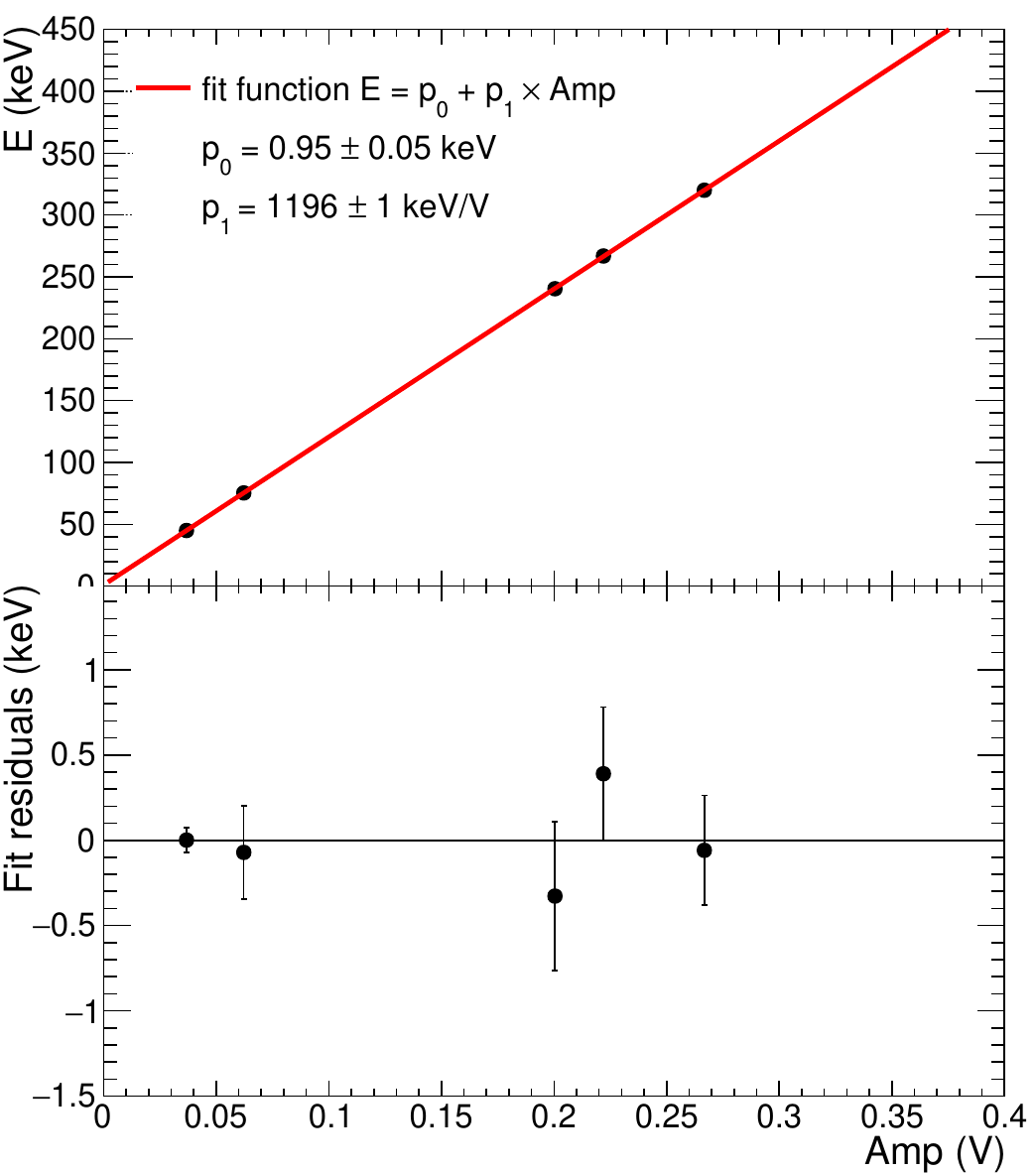}
    \caption{(top) Calibration curve of kinetic energy E (keV) {\it vs} signal amplitude (V) obtained with conversion electrons from a \Ba\ source. (bottom) Difference between the fitted value and the expected energy of the conversion electron.}
    \label{fig:fig4}
\end{figure}

The energy calibration of the detector is performed using conversion electron of \Ba\ source at 45.01, 75.38, 240.41, 266.87 and 320.03\,keV. The distribution of the maximum  of the pulse height value measured in the detector, for a given fixed value of the magnetic field, is used to obtain the energy of the incoming electrons using the mean value of a Gaussian fit (Fig.~\ref{fig:fig3}). A fixed magnetic field value is set which selects a fixed electron energy. The maximum of the pulse height measured in the detector corresponds to the deposited energy by the focalized electron. The distribution of these maximum amplitudes is shown in Fig.~\ref{fig:fig3} for electrons energy of a \Ba\ source at 45.01\,keV and 75.38\,keV. No asymmetry is observed in the peaks, which corresponds to the fully deposited energy, therefore, a Gaussian fit is performed and its obtained mean value is used for the energy calibration. This value for each electron energy of the \Ba\ source is used to obtain the energy calibration curve of the kinetic energy~(E) as a function of  the maximum amplitude of the pulse height~(Amp) (Fig.~\ref{fig:fig4}). To suppress distortion effects due to the hysteresis of the magnetic field, the energy corresponding to each measured point in the spectrum is obtained using the mean of a Gaussian fit of the amplitude of the pulse height distribution and no longer using the value of the current in the coils as performed in the previous measurements~\cite{Juget2014-il,Juget2019-fj}. This allows to have an online calibration and take into account possible deviations of the magnetic field for a given current.

\begin{figure}[!h]
    \centering
    \includegraphics[width=0.65\textwidth]{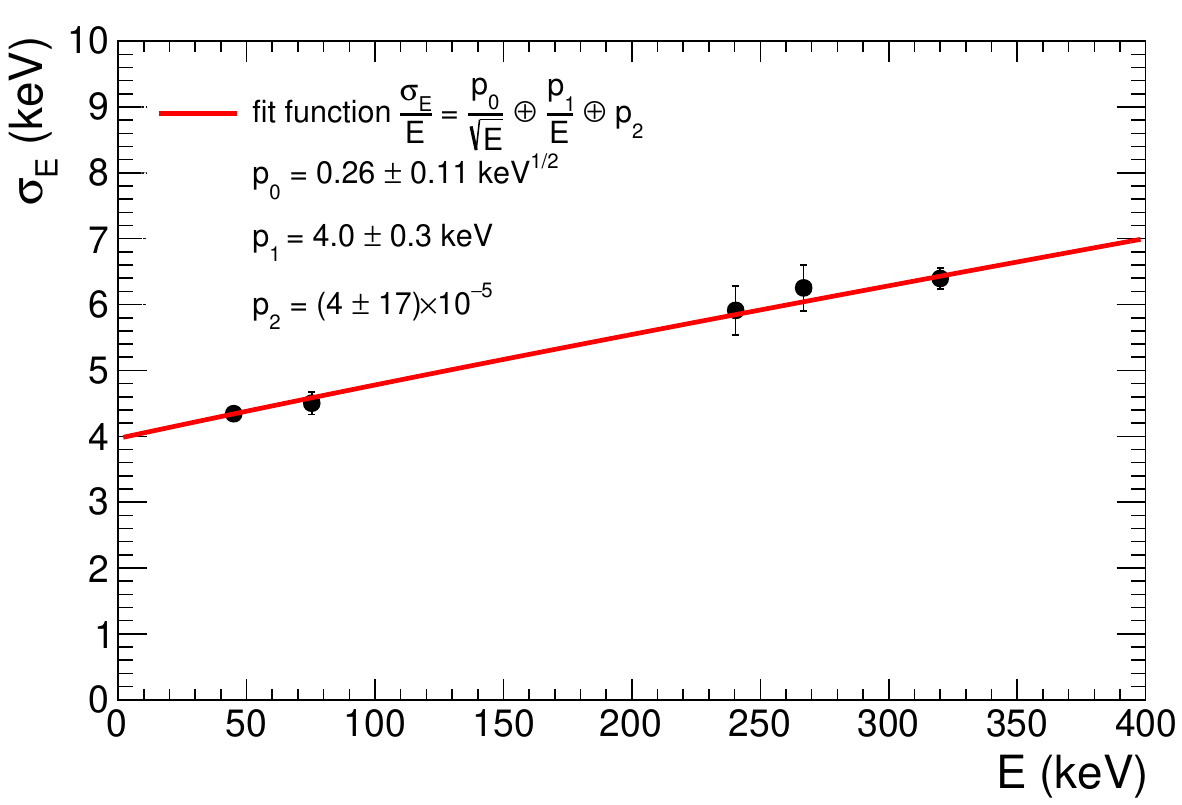}
    \caption{Energy resolution ($\sigma_{\rm E}$) measured with \Ba\ conversion electrons with the parameter values obtained from the fit.}
    \label{fig:fig5}
\end{figure}

The energy resolution was obtained using the conversion electrons from the \Ba\ source and the width of the Gaussian fit on each peak. The energy resolution as a function of the energy is fitted using the following relation: 
$\frac{\sigma_{\rm E}}{E}=\frac{p_0}{\sqrt{E}}\oplus\frac{p_1}{E}\oplus{p_2}$, 
where $p_0$, $p_1$ and $p_2$ are fit parameters. The result is shown in Fig.~\ref{fig:fig5}. The last parameter $p_2$ could have been omitted in the fit as it has a large uncertainty and is compatible with 0. Usually this parameter corresponds to the imperfections of the calorimeter (detector) construction, the non-uniformity of the detector response, the fluctuation in energy containment or the energy lost in dead material. As the resolution is energy-dependent, an algorithm based on the method detailed in Ref.~\cite{Wortman1964-ar}, is applied to correct the measured spectra in the offline analysis. It uses an iterative technique to avoid magnification of statistical fluctuation by performing the correction on an n$^{\rm th}$-order polynomial fitted to the experimental data.

\subsection{Efficiency calculation}
\label{sec:efficiency}
\begin{figure}[!h]
    \centering
    \includegraphics[width=0.85\textwidth]{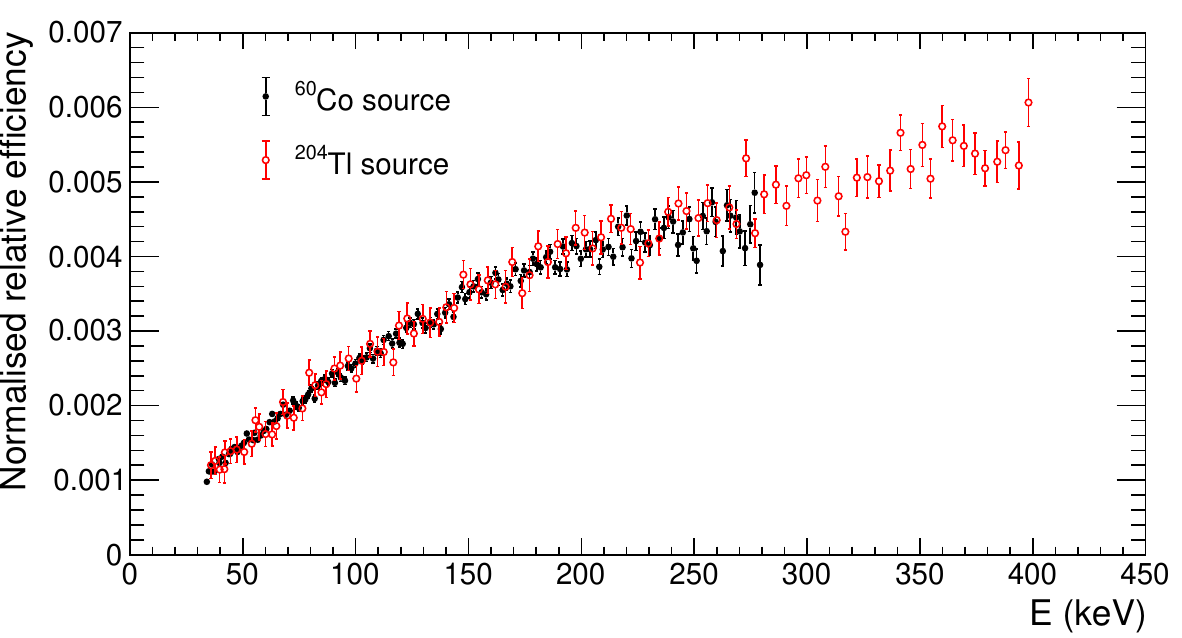}
    \caption{Efficiency versus kinetic energy obtained with \Co\ (black points) and \Tl\ (open points) spectra.}
    \label{fig:fig6}
\end{figure}

\begin{figure}[!h]
    \centering
    \includegraphics[width=0.85\textwidth]{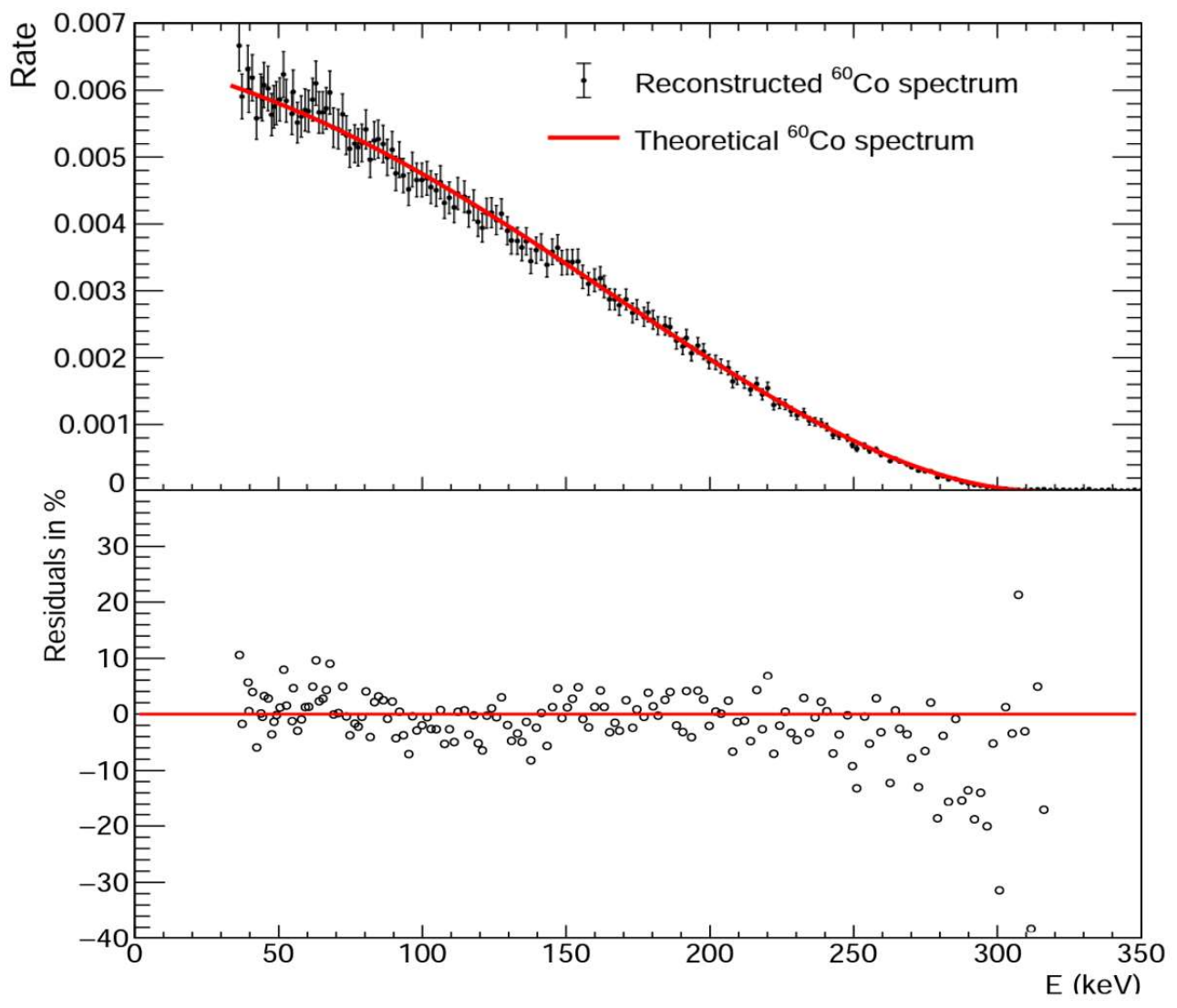}
    \caption{Beta spectrum of \Co\ (black points) reconstructed using the efficiency obtained with the \Tl\ spectrum and comparison with the expected theoretical spectrum calculated with the BetaShape software(top). Residuals in \% (Bottom) where the red line represents the 0 value.}
    \label{fig:fig7}
\end{figure}

The relative efficiency of the spectrometer was calculated using a \Co\ source for which the energy spectrum can be precisely calculated using the BetaShape software~\cite{Mougeot:2015bva}. The ratio between the measured spectrum and the theoretical spectrum gives the relative efficiency versus the energy. Additionally the spectrum of a \Tl\ source was measured to validate the efficiency calculation. Both theoretical spectra are calculated using the BetaShape software. Figure~\ref{fig:fig6} shows the comparison between the two efficiencies and Fig.~\ref{fig:fig7} gives the \Co\ beta spectrum reconstructed with the efficiency calculated using the \Tl\ spectrum, as well as the comparison with the theoretical spectrum.

\newpage
\newpage
\section{Source preparation}
\label{sec:source}
\Tm\ was produced by thermal neutron irradiation of \isotope[170]{Er}$_2$O$_3$ in V4 beam tube of the high-flux research reactor at Institut Laue-Langevin, France (ILL). The V4 beam tube is a vertical beam tube protruding into the heavy water tank of the ILL reactor. The \isotope[170]{Er} sample had been irradiated at about 20 cm distance from the fuel element. There the neutron spectrum is dominantly thermal, i.e. $>$90\% of all neutrons are in the thermal regime and $>$97\% of all neutrons have energies $<$ 1 eV ~\cite{Köster:2012,Letourneau:2005}. \isotope[171]{Tm} was chemically separated from the starting material as well as other chemical impurities at the Paul Scherrer Institute, Switzerland.  In particular, a two-step chromatography method was implemented. In the first step, the lanthanides were fractionated using the cation exchange resin AMINEX (Bio-Rad, USA), via gradient elution with the chelating agent $\alpha$-Hydroxyisobutyric acid ($\alpha$-HIBA). In the second step, all the fractions containing \Tm\ were unified and acidified with 1\,M HNO$_3$, then loaded into a column containing the lanthanide-specific LN-resin to remove the chelating agent $\alpha$-HIBA. A complete elution of pure \Tm\ was achieved using 3\,M HNO$_3$. The \isotope[170]{Tm}/\isotope[171]{Tm} activity ratio was reported to equal 3.3.10$^{-3}$ on November 01, 2014.    More details about this chemical purification and the characterization of \Tm\ can be found in ~\cite{Heinitz2017-lg}.

 A source was prepared from a solution remaining from that work by drop deposition and drying on a support as described in ~\cite{Juget2019-fj}, the activity of the source being 50(3) kBq at the begining of the measurement in spring 2022. 
 The \isotope[170]{Tm}/\isotope[171]{Tm} activity ratio has reduced to 2.2.10$^{-8}$ due to the shorter half-life of \isotope[170]{Tm} (T$_{1/2}$= 128.6(3) days) compared to \isotope[171]{Tm} (T$_{1/2}$=1.92(1) years). This $"$pre-aging$"$ of an already highly enriched \isotope[171]{Tm} source assures that no interference from beta decay of \isotope[170]{Tm} possibly present as impurity. Therefore, the $^{170}$Tm impurity could be neglected in the underlying measurement.

\section{\Tm\ Measurement}
\label{sec:measurement}
\subsection{Spectrum measurement}
\label{sec:spectrum}
\begin{figure}[!h]
    \centering
    \includegraphics[width=0.85\textwidth]{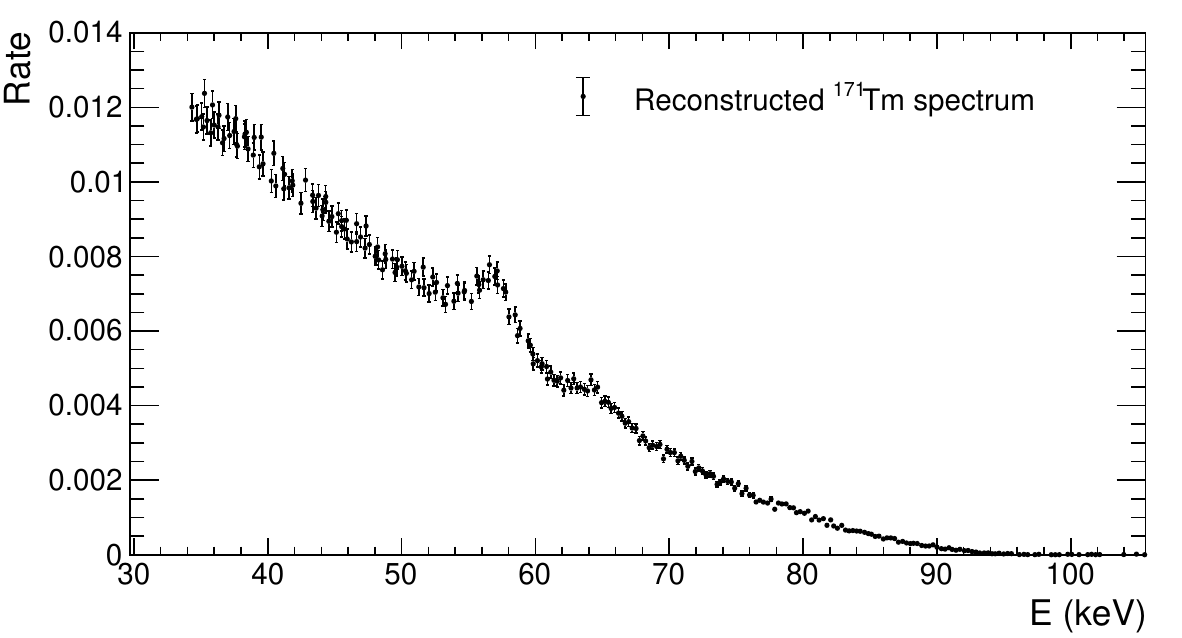}
    \caption{Measured spectrum of \Tm\ with only the statistical uncertainty for the rate.}
    \label{fig:fig8}
\end{figure}

\begin{figure}[!h]
    \centering
    \includegraphics[width=0.85\textwidth]{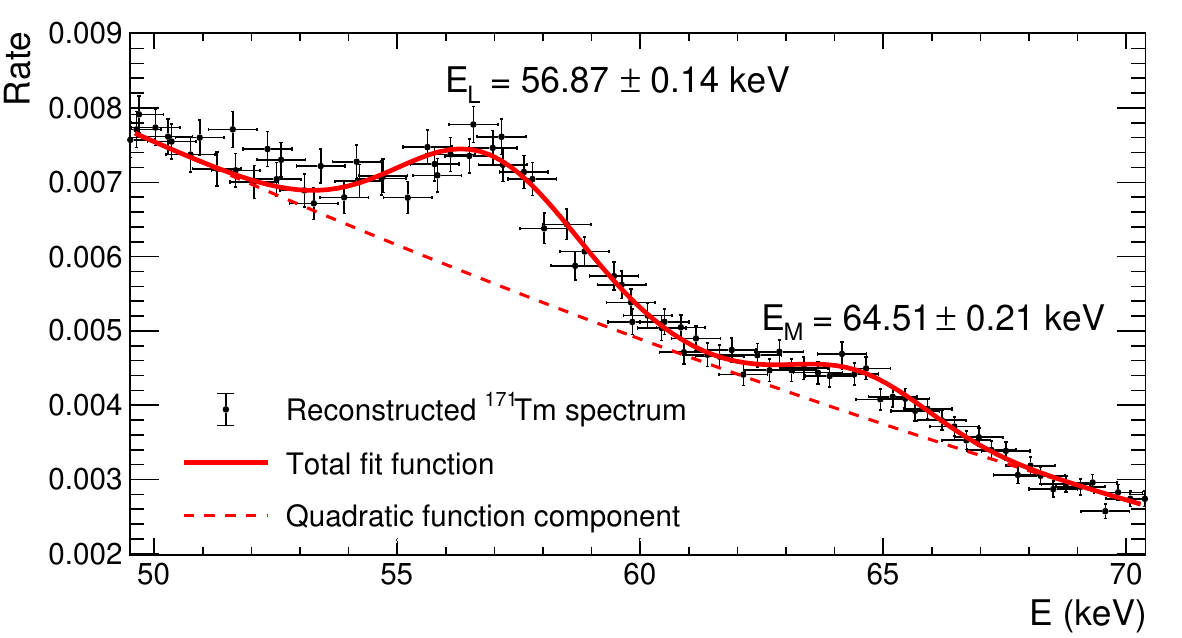}
    \caption{Gaussian fit of the two conversion electrons peaks. The mean value of the Gaussian fit with its uncertainty is shown. The obtained values, 56.87(14) and 64.51(21)\,keV, respectively, have to be compared with the expected energy values 57.06(72)\,keV and 64.97(38)\,keV.}
    \label{fig:fig9}
\end{figure}

By increasing the current of the magnet, the spectrum shape is measured from 34\,keV to 108\,keV in 0.25\,keV steps,  each measurement lasting 72 minutes. The background is estimated by performing the same measurement point by point, without source. The obtained background count rate for each point is removed from the data obtained with the source. Figure~\ref{fig:fig8} gives the obtained spectrum after reconstruction with the efficiency obtained with the \Co\ and corrected for the energy resolution. A Gaussian fit on the two resolved conversion electron peaks gives respectively the mean energy 56.87(14) and 64.51(21)\,keV (Fig.~\ref{fig:fig9}), where the uncertainty is the one from the fit only. These values have to be compared with the expected energy between 56.26-57.79\,keV (L lines with intensity 0.66(5)\%) for the first peak and between 64.33-66.72\,keV (MNOP lines  with intensity 0.2024(12)\%) for the second peak~\cite{ENSDF}. The calculation of the intensity-weighted average energies with {\it BrIcc}~\cite{Kibedi:2008ydo}, using the mixing ratio of $\delta$=+0.684(17)~\cite{ENSDF}, gives 57.06(72)\,keV for the first peak and 64.97(38)\,keV for the second peak. The experimentally determined intensity ratio of the L/(M+N+O+P) is 3.68(79), to be compared with the calculated ratio of 3.29(7) using {\it BrIcc}.

\subsection{Maximum energy determination}
\label{sec:emax}
The beta spectrum is usually described by
\begin{equation}
    N(W)\dd{W}=K\cdot F(W,Z)\cdot p\cdot W\cdot(W_0-W)^2\cdot C(W)\cdot X(W)\cdot r(Z,W)\dd{W},
    \label{eq:beta}
\end{equation}
where the energy $W=1+\frac{E}{m}$, $N(W)$ is the measured rate, $K$ a constant, $F(Z,W)$ the Fermi function, $p=(W^2-1)^{1/2}$ the momentum, $W_0 = 1+\frac{E_{\rm max}}{m}$ ($E_{\rm max}$ being the end point energy of the electron). The additional correction functions, $C(W)$ is the shape factor, which is equal to 1 in the allowed transition, $X(W)$ allows for screening and atomic exchange effects and $r(Z,W)$ accounts for the atomic overlap effect. 

In the case of \Tm, considering the $\xi$-approximation, the shape factor correction is taken as $C(W) = 1$. The $X(W)$ correction was provided directly by private communication with Xavier Mougeot and was calculated as described in Refs.~\cite{Mougeot:2015bva,XENON:2020rca,Haselschwardt:2020iey}. The $r(Z,W)$ correction was calculated as described in Ref.~\cite{Hayen:2017pwg} and can be written as $r(Z,W) = 1 - \frac{1}{W_0-W}\cdot B(G)$, where $B(G)$ is a constant which is calculated using the parametrization from Ref.~\cite{Haselschwardt:2020iey}, $B(G) = 0.286155$. 

\begin{figure}[!h]
    \centering
    \includegraphics[width=0.85\textwidth]{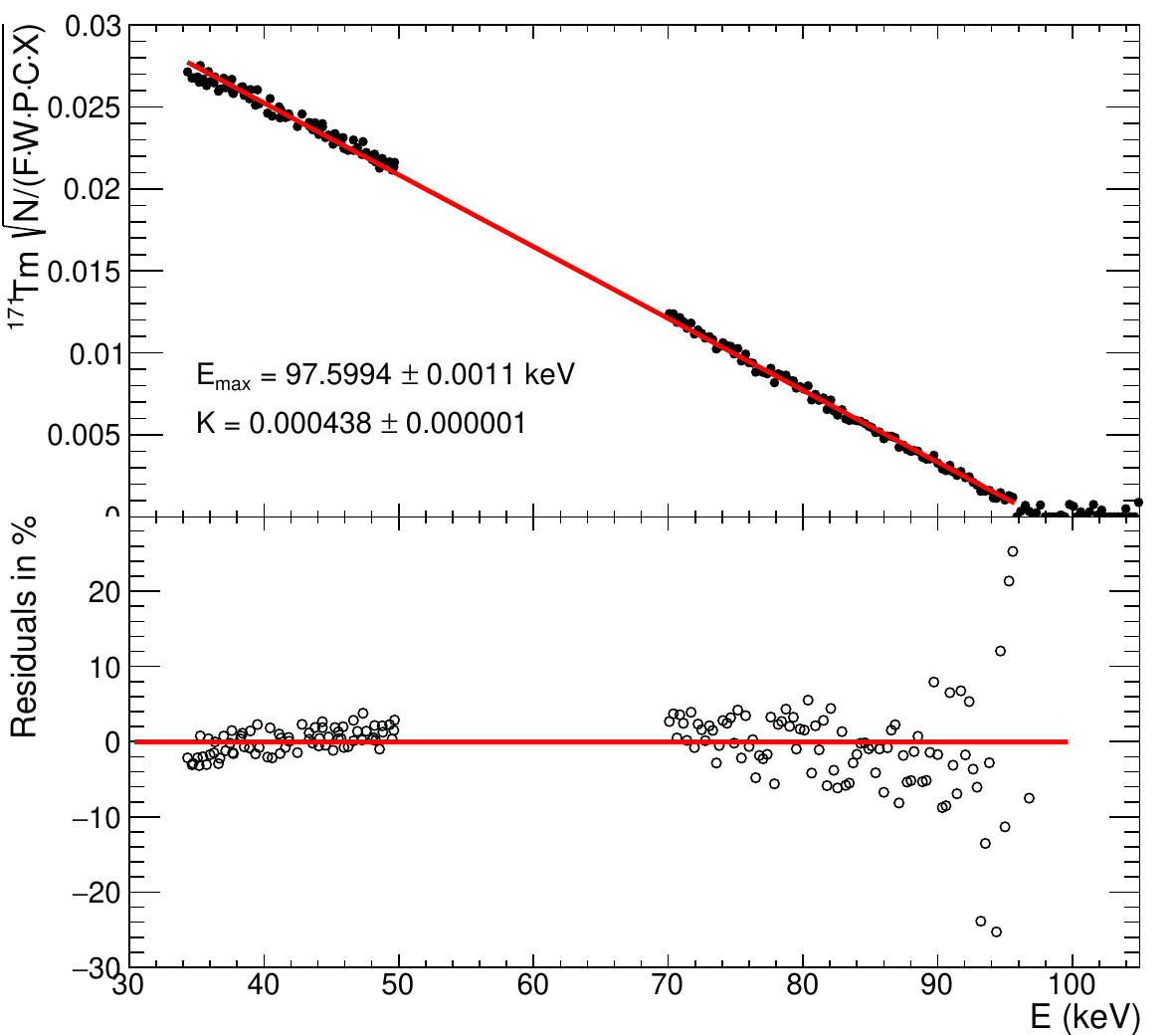}
    \caption{Kurie plot (top) using the Eq.~\ref{eq:beta1} with a fit from 34 keV to 96 keV. The uncertainty corresponds to the fit only. Residuals of the Kurie plot (bottom), where the line represents the 0 value.}
    \label{fig:fig10}
\end{figure}

Following the Kurie plot procedure~\cite{Kurie1936-qy}, Eq.~\ref{eq:beta} can be rewritten as~\cite{Kossert:2022ydo}
\begin{equation}
    \sqrt{\frac{N(W)}{F(W,Z)\cdot p\cdot W\cdot C(W)\cdot X(W)}} = K\sqrt{(W_0-W)^2-0.286155\cdot(W_0-W)},
    \label{eq:beta1}
\end{equation}
where $K$ and $W_0$ can be obtained with a fit procedure and therefore infer the value of $E_{\rm max}$.
Figure~\ref{fig:fig10} gives the Kurie plot with the obtained fit values in the energy range 34-96\,keV where the middle part from 50 to 70\,keV containing the conversion electron peaks has been removed. The end point energy is \Emax = 97.60 keV. The residual difference between the fitted curve values and the measured points is also shown in Fig.~\ref{fig:fig10}. The agreement is well observed over the whole range taking into account the measurement fluctuations. 

In order to estimate the uncertainty due to the energy range, several fits were performed where each limit was extended or restricted to 41\,keV for the low limit and reduced to 88\,keV for the high limit in 1\,keV steps. The resulting values ranged from 97.34 to 97.83\,keV and therefore an uncertainty of 0.26\,keV (difference between \Emax = 97.60\,keV and 97.34\,keV) was associated to the \Emax\ value for the fit method.

Data points above 98\,keV were used to estimate the zero-line fluctuation due to background subtraction. Their standard deviation was used to calculate the uncertainty of the zero-line estimation. The \Emax\ value was estimated by removing and adding this standard deviation and an uncertainty of 0.18\,keV was associated to the \Emax\ value for the background fluctuation. To estimate the uncertainty coming from beta spectrum model, the fit was performed by removing the exchange effect correction and the atomic overlap effect. The \Emax\ value found was 97.51\,keV, therefore an uncertainty of 0.09\,keV was associated to the beta spectrum model.

Concerning the other uncertainty components, energy calibration, statistics and energy resolution, a Monte Carlo simulation method was used according to the ``Guide to the expression of uncertainty in measurement''~\cite{Cox2005-gg}. It consists of varying each quantity within its uncertainty 10000 times and redoing the fit each time to obtain the effect variation of the \Emax\ value. The distribution of \Emax\ values gives a Gaussian shape and its width obtained by a fit procedure is taken as the uncertainty for the component under consideration. Table~\ref{tab:Uncertanty} gives the obtained values of each uncertainty component. For the final value of the uncertainty, the different components are assumed uncorrelated and are summed in quadrature.  The final value of \Emax\ is 97.60 $\pm$ 0.38\,keV.

\newpage
\section{Discussion}
\label{sec:discussion}
The improvements realized on the spectrometer have reduced the low energy threshold down to 34\,keV. The Kurie plot performed over a large energy range for the \Tm\ main transition gives an \Emax\ value of 97.60(38)\,keV, which is in agreement within the standard uncertainty with the two last published measurements (Fig.~\ref{fig:fig2}), but in disagreement with the measurement of 1955 ~\cite{Bisi}. For this measurement the article gives no explanation about the uncertainty evaluation and no estimation of the impurity amount is given in the article. The presence of \isotope[170]{Tm} impurity, which still was reduced by waiting its decay after fourteen months (\isotope[170]{Tm} half-life is 128.6(3) days compare to 1.92(1) years for \isotope[171]{Tm}), could explain the observed difference for the Emax value. For the 1953 publication ~\cite{Hollander1953-zr}, they refer to measurement reported in Oak Ridge National Laboratory Report ORNL-65 (July 1948) which still remains classified.

\begin{figure}[!h]
    \centering
    \includegraphics[width=0.85\textwidth]{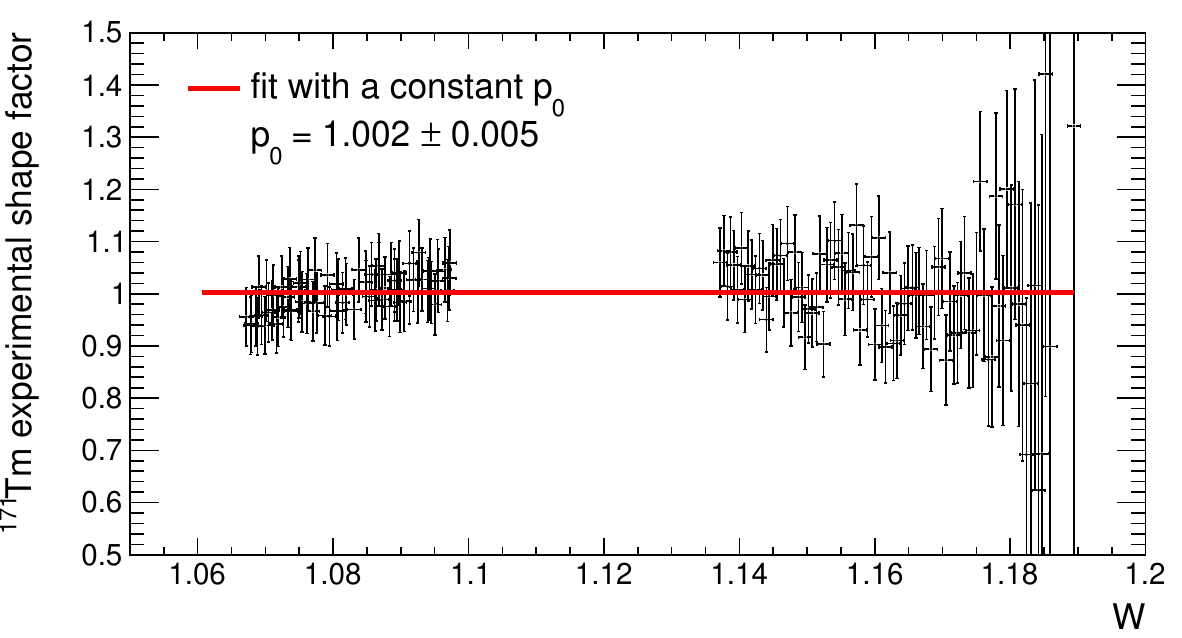}
    \caption{Experimental shape factor versus $W$, where a constant fit gives a value close to 1.}
    \label{fig:fig11}
\end{figure}

In the Kurie plot (Fig.~\ref{fig:fig10}) the data are well fitted confirming that the spectrum is compatible with the allowed transition. The experimental shape factor has been assessed using 
$\frac{N(W)}{F(W,Z)\cdot p\cdot W\cdot X(W)\cdot r(Z,W)}$. The result, shown in Fig.~\ref{fig:fig11}, is in agreement with a flat line, and a constant fit over the whole energy range gives a value close to 1.
This confirms that the $\xi$-approximation can be used to calculate the beta spectrum as there is no energy dependence of the shape factor. Therefore the beta spectrum of the \Tm\ has the same shape as an allowed decay up to O(1\%) corrections.
\begin{table}[!ht]
    \centering
    \caption{Uncertanty budget for \Emax\ value at k=1.}
    \begin{tabular}{l c c l}\\\hline
    Uncertainty components & Unc.  & Rel. unc. & Comments \\
    &(keV) &  (\%) & \\ \hline
    Fit range and method	&0.26 &	0.27	& Variation of the energy range \\
    & & & and uncertainty from the fit \\
    & & & method\\
    Background	& 0.18	& 0.18	& Zero-line fluctuation \\
    Energy calibration &	0.04	& 0.04 &	Uncertainty from \Ba\ \\
    & & &  conversion electrons  \\
    & & &  measurement \\
    Statistics	& 0.12& 	0.12& 	Statistics + efficiency - dead-time \\
    Energy resolution (method) &	0.14 &	0.14	& Resolution uncertainty from \\
    & & & \Ba\ conversion electrons peaks \\
    & & & measurement and method used  \\
    & & & for deconvolution \\
    Model	& 0.09	& 0.09	& Without Exchange effect and \\
    & & & atomic overlap \\
    {\bf Combined uncertainty} & 	{\bf 0.38}	& {\bf 0.39}	& \\\hline
    \Emax (keV) & 97.60(38) & &\\\hline  
    \end{tabular}
    \label{tab:Uncertanty}
\end{table}
\smallskip

\newpage
\bigskip
\newpage
\section*{Acknowledgments}
\label{sec:acknowledgment}
We acknowledge Xavier Mougeot from LNHB, for his advice and calculations of the theoretical part of this work. The spectrometer was operated and maintained thanks to the EPFL workshop team. This project has received funding from EPFL School of Basic Sciences. The work of Alexey Boyarsky was supported by SNSF Scientific Exchanges grant 207121.

\bibliographystyle{JHEP}
\bibliography{references}

\end{document}